\documentclass[cernpreprint, atlasdraft=false, texlive=2023, USenglish, texmf, orcidlogo]{atlasdoc}
\usepackage{atlaspackage}
\usepackage{atlasbiblatex}

\usepackage[hion=true]{atlasphysics}
\usepackage{multirow}
\usepackage{dcolumn}
\usepackage[export]{adjustbox}
\usepackage{verbatim}
\usepackage{physics}

\graphicspath{{logos/}{figures/}}

\usepackage{ANA-HION-2022-19-PAPER-defs}
\usepackage{mathrsfs}
\usepackage{enumitem}
\AtlasTitle{Azimuthal angle correlations of muons produced via
heavy-flavor decays in 5.02~\TeV\ Pb+Pb and \pp\ collisions with the ATLAS
detector}

\AtlasAbstract{%
Angular correlations between heavy-quarks provide a unique probe of the
quark-gluon plasma created in ultra-relativistic heavy-ion collisions.
Results are presented of a measurement of the azimuthal angle correlations
between muons originating from semileptonic decays of heavy-quarks
produced in 5.02~\TeV\ Pb+Pb and \pp\ collisions at the LHC\@.
The muons are measured with transverse momenta and pseudorapidities
satisfying $p_{\mathrm{T}}^\mu > 4$~\GeV\ and $|\eta^\mu|<2.4$, respectively.
The distributions of azimuthal angle separation, $\Delta\phi$, for muon pairs
having pseudorapidity separation $|\Delta\eta|>0.8$, are measured in
different Pb+Pb centrality intervals and compared to the same distribution
measured in \pp\ collisions at the same center-of-mass energy.
Results are presented separately for muon pairs with opposite-sign charges,
same-sign charges, and all pairs.
A clear peak is observed in all $\Delta\phi$ distributions at
$\Delta\phi\sim\pi$, consistent with the parent heavy-quark pairs being
produced via hard-scattering processes.
The widths of that peak, characterized using Cauchy-Lorentz fits to the
$\Delta\phi$ distributions, are found to not vary significantly
as a function of Pb+Pb collision centrality and are similar for \pp\
and Pb+Pb collisions.
This observation will provide important constraints on theoretical descriptions
of heavy-quark interactions with the quark-gluon plasma.
}

\PreprintIdNumber{CERN-EP-2023-185}

\AtlasJournalRef{\PRL 132 (2024) 202301}
\AtlasDOI{10.1103/PhysRevLett.132.202301}

\hypersetup{pdftitle={ATLAS document},pdfauthor={The ATLAS Collaboration}}

\begin{document}
\maketitle

Heavy-flavor (HF) quarks have long been considered important probes of the
quark-gluon plasma (QGP) created in ultra-relativistic heavy-ion
collisions~\cite{Braaten:1991we,vanHees:2005wb,vanHees:2007me,Rapp:2008qc,Rapp:2009my,Gossiaux:2009hr,Berrehrah:2014kba,Andronic:2015wma,Dong:2019unq,Dong:2019byy,}.
They are %
produced primarily in hard-scattering
processes that occur at early times in the nuclear collisions with
the result that heavy-quarks experience the full evolution of the QGP.
The interactions of the heavy-quarks with the QGP can induce energy
loss or ``quenching'' that is experimentally observed in the
suppression of the yield of HF hadrons~\cite{STAR:2014wif,ALICE:2015ccw,ALICE:2015vxz,CMS-HIN-16-001,CMS-HIN-16-011}
and their decay products~\cite{PHENIX:2005nhb,PHENIX:2010xji,PHENIX:2014rwj,HION-2015-06,ALICE:2016uid,HION-2019-58,CMS-HIN-10-006,CMS-HIN-14-005,HION-2016-07,CMS-HIN-16-016},
or heavy-quark jets, especially bottom-tagged jets~\cite{CMS-HIN-12-003,HION-2018-24}.
Separately, the coupling of the heavy-quarks to collective expansion
of the plasma~\cite{Nahrgang:2014vza,Hippert:2019swu,},
can be observed via
the azimuthal anisotropy of HF hadrons~\cite{ALICE:2013olq,ALICE:2014qvj,ALICE:2020iug,CMS-HIN-16-007}
or their decay products~\cite{PHENIX:2010xji,PHENIX:2014rwj,HION-2017-05}.

Measurements sensitive to the angular deflection of heavy quarks in
the plasma may provide an important, alternative probe of quark
interactions in the plasma.
On very general physics grounds, the
interactions between the quarks and the hot plasma  would be expected
to induce some deflection of the quarks. Indeed, at momenta comparable
to or less than the quark mass, HF quarks
are thought to undergo Brownian motion
with their transport being characterized by a
diffusion coefficient~\cite{Casalderrey-Solana:2006fio,}
(also see Ref.~\cite{Andronic:2015wma} and references therein).
At momenta much higher than the quark mass, the quarks would,
in a weak-coupling scenario~\cite{Djordjevic:2003zk,Wicks:2007am}, be expected to multiply scatter
in the plasma and thereby be  ``collisionally'' broadened. However,
the scattering will also stimulate gluon
emission~\cite{Djordjevic:2003zk,Moore:2004tg,Romatschke:2004au,Djordjevic:2004nq,Wicks:2007am,Peigne:2008nd,Gossiaux:2010yx,Djordjevic:2013xoa,}
that may damp the collisional broadening. The relative importance of
the multiple scattering and radiative damping on the propagation of
the heavy quarks, especially bottom quarks, is not currently well-constrained by data.

The angular deflections of the quarks can be experimentally probed by measuring
azimuthal angles between quarks and anti-quarks created in the same
hard-scattering process, where momentum conservation induces a strong angular correlation
between the two particles. Indeed, theoretical calculations have shown
that angular correlations of bottom quarks are very sensitive to the relative
importance of collisional and radiative scattering
processes~\cite{Nahrgang:2013saa} in the bottom quark-medium interaction.
The direct detection of bottom hadron pairs is
difficult due to their complex decays, but detection of pairs of
leptons resulting from simultaneous semileptonic decays of B hadrons
is experimentally feasible, though  no such measurements have
been made prior to this letter.

At LHC collision energies,
muons having transverse momenta greater than a few \GeV\ predominantly
result from, HF decays~\cite{STDM-2011-07}.
Both charm and bottom hadrons contribute to single-muon production,
but the contribution of charm quarks to
the production of muon pairs where both muons have $\pT > 4$~\GeV, was found to be kinematically
suppressed relative to bottom quarks, by about an order of
magnitude in a
\powheg~\cite{Nason:2004rx,Frixione:2007vw,Alioli:2010xd,Sjostrand:2007gs,Sjostrand:2014zea} study in Ref.~\cite{SupplementaryMaterial}.
The charges of the muons provide a further handle to suppress
the contributions from charm.
Namely, pairs of charm hadrons produced in hard-scattering processes will generate,
nearly exclusively, opposite-sign muon pairs.
However, in decays of B hadrons, muons may be produced either from the bottom
hadrons themselves or from the decay of the secondary charm hadrons,
in which case, the sign of the resulting muon can be reversed.
As a result, \bbbar\ pairs can produce two muons of the same sign.
Separately, the mixing of neutral B mesons provides an
efficient mechanism for generating same-sign muon pairs.
Due to the short oscillation time (compared to the decay time)
of neutral $B$ mesons~\cite{LHCb:2021moh}, about half of the muon pairs produced
when one or both parents is a neutral $B$ meson are of the same sign.
Thus, same-sign muon pairs provide a clean probe of \bbbar\ production, and
a comparison of measurements using opposite-sign and same-sign
muons allows potential contributions from \ccbar\  pairs to
be estimated.

This Letter presents ATLAS measurements of angular correlations between muons in
opposite- and same-sign pairs produced in both \PbPb\ and \pp\
collisions at $\sqn = 5.02$~\TeV\ using datasets with integrated
luminosities of 1.94~\inb\ and 0.26~\ifb, respectively.
The muons, which primarily result from semileptonic decays of charm and
bottom hadrons, are measured over the transverse momentum ($\pT^\mu$)
and pseudorapidity~\cite{AtlasCoordinate} ($\eta^\mu$) ranges,
$\pT^\mu > 4$~\GeV\ and $|\eta^\mu|< 2.4$.
Additionally, the average \pt\ of the two muons in the pair, \ptbar,
is required to be larger than 5~\GeV.
The $\ptbar>5~\GeV$ requirement removes pairs
with very small trigger and reconstruction efficiencies (described below)
and improves the statistical precision of the measurements.
To suppress contributions from quarkonia decays and from the production of
heavy-quark pairs within jets, the two muons are required to have a pseudorapidity
separation, $\absDy = |\eta^{\mu_1} - \eta^{\mu_2}| > 0.8$ where $\mu_1$ and $\mu_2$ refer to the
two muons in the pair with arbitrary ordering.
Distributions of azimuthal angle separation, $\Dphi = \phi^{\mu_1} - \phi^{\mu_2}$,
are measured in different \PbPb\ collision centrality~\cite{Miller:2007ri}
intervals and in \pp\ collisions.
Backgrounds from Drell–Yan (DY) processes are estimated using Monte Carlo
(MC) simulations, and are subtracted from the \Dphi\ distributions.
Results are obtained for same-sign, opposite-sign, and all pairs,
all of which show clear enhancement on the ``away-side'' (i.e., at $\Dphi = \pi$),
consistent with contributions from hard-scattering processes.
The widths of the away-side peaks are characterized
by fitting the \Dphi\ distributions with Cauchy-Lorentz functions
which describe the data well.
The centrality-dependence of the width %
is the focus of the measurements in this paper.

The ATLAS experiment~\cite{PERF-2007-01} at the LHC is a multipurpose particle detector
with a forward--backward symmetric cylindrical geometry and a
near \(4\pi\) coverage in solid angle.
It consists of an inner tracking detector (ID) surrounded by a thin superconducting solenoid
providing a \SI{2}{\tesla} axial magnetic field, electromagnetic and hadronic calorimeters, and a muon spectrometer.
The ID covers the pseudorapidity range \(|\eta| < 2.5\).
The calorimeter system consists of a liquid-argon (LAr) electromagnetic
calorimeter covering $|\eta|<3.2$, a steel/scintillator sampling
hadronic calorimeter covering $|\eta|<1.7$, a LAr hadronic calorimeter
covering $1.5 < |\eta| < 3.2$, and a LAr forward calorimeter (FCal)
covering $3.1 < |\eta| <4.9$.  The muon spectrometer surrounds the
calorimeters and is based on three large superconducting air-core
toroidal magnets with eight coils each.
It includes a system of precision tracking chambers and fast detectors
for triggering.
A two-level trigger system~\cite{TRIG-2016-01}, the first (L1) implemented in hardware and
the second (HLT) implemented in software, is used to select events
for this measurement.
An extensive software suite~\cite{ATL-SOFT-PUB-2021-001} is used in the
reconstruction and analysis of real and simulated data, in detector
operations, and in the trigger and data acquisition systems of the experiment.

The \PbPb\ events are classified into centrality percentiles
based on the total transverse energy measured in the
FCal~\cite{HION-2016-06,HION-2017-10}.
In this classification, the 0--10\% centrality interval
corresponds to the 10\% of inelastic \PbPb\ collisions
with the smallest impact parameter~\cite{Miller:2007ri}.
Events selected by two dimuon triggers are used in this analysis.
The first trigger required a single muon with $\pt>4~\GeV$ at L1
and two muons having $\pT > 4$~\GeV\ at the HLT\@.
The second trigger required two muons with $\pt>4~\GeV$ at both L1 and the HLT\@.
Both triggers were used in the \PbPb data,
and are further described in Ref.~\cite{HION-2020-10}.
In the \pp\ data taking,  the first trigger was not implemented and all
data used here were recorded by the second trigger.

Events used in the analysis are required to be recorded during stable
running conditions of the LHC, to have no detector hardware or readout
error, and to have a reconstructed collision vertex.
Charged-particle tracks and collision vertices are reconstructed from
hits in the ID using standard methods~\cite{PERF-2015-08}.
Muons are reconstructed by combining ID tracks with tracks
reconstructed in the muon spectrometer.
The muons are required to satisfy the \textit{tight}
muon selection requirements~\cite{PERF-2015-10},
which reduce the fraction of pairs containing a muon resulting from
pion or kaon decay to less than 5\% across the measured centrality range
(estimated via the procedure detailed in Ref.~\cite{HION-2015-06}).
For events containing more than two muons that satisfy the selection
and kinematic requirements (about 2\% of all events),
all possible pairs are considered.

As noted above, to suppress contributions from quarkonium decays, muon
pairs used in the measurement are required to have $\absDy > 0.8$.
This selection essentially eliminates dimuons produced by light vector mesons and
\jpsi decays, but still admits a
contribution from $\Upsilon$ decays.
To remove those as well as pairs from $Z$-decays,
opposite-sign pairs having masses in the ranges
$[9.2,10.4]$~\GeV\ and $[70,110]$~\GeV\ are excluded from the measurement.
In the \PbPb data, dimuons produced in $\gamgam\rightarrow \mu^+\mu^-$
scattering processes~\cite{HION-2018-11,HION-2020-10,HION-2016-02},
are almost fully removed by requiring the opposite-sign muon pairs to have
$|\Dphi-\pi|/\pi>0.01$ or $|\pT^{\mu_1}-\pT^{\mu_2}|/(\pT^{\mu_1}+\pT^{\mu_2})>0.08$.

To account for the loss of muon pairs due to reconstruction and
trigger inefficiencies, each pair is assigned a weight
which is the inverse of the product of the reconstruction and
trigger efficiencies for the pair~\cite{HION-2020-10}.
The efficiencies are calculated as a function of the \pt\ and $\eta$
of the two muons in the pair~\cite{HION-2020-10}.
The average weight for the pairs is $\sim2.3$ ($\sim2.4$) in the \pp\ (\PbPb) data set.
A separate acceptance correction is applied to opposite-sign pairs to
account for the losses resulting from the dimuon mass requirements
to reject $\Upsilon$ and $Z$-bosons,
and the requirements applied to suppress
$\gamgam\rightarrow \mu^+\mu^-$ events.
These are obtained by applying the mass and \gamgam\ requirements
to the same-sign pairs and evaluating the \Dphi-dependent fraction of
pairs that survive.
That fraction is taken to be the acceptance for opposite-sign pairs,
$A^{\text{opp}}(\Dphi)$, and is used to correct the measured
distribution for the opposite-sign pairs.
To check the sensitivity of the acceptance to possible differences in the
single-muon and pair kinematics between same-sign and opposite-sign pairs,
separate estimates of the pair acceptance are obtained using \textit{mixed-events}.
In the mixed-event estimate, opposite-sign muon pairs are made using muons reconstructed
in separate events, and $A^{\text{opp}}(\Dphi)$ is estimated by evaluating
what fraction of pairs satisfy the mass and \gamgam\ requirements.
The events used to construct the mixed-event distributions only require a single muon
in each event, as the pair is constructed by combining two events.
Therefore, in order to have more statistics in the mixed-event distributions,
the events used to make the mixed-event distributions are selected using
a trigger that only required a single muon with $\pt>4~\GeV$ at both L1 and the HLT\@.
Differences between the two estimates of $A^{\text{opp}}(\Dphi)$ affect
the final observables by less than 0.5\%, and are included
as systematic uncertainties.
Contributions to the \Dphi\ distributions from DY
processes are evaluated using a
\powheg~\cite{Nason:2004rx,Frixione:2007vw,Alioli:2010xd,Sjostrand:2007gs}
MC setup, further described in Ref.~\cite{HION-2020-10}.
The estimated DY contribution is then subtracted from the measured distributions.

The measurements of muon angular correlations are presented in the
form of two-muon correlation functions:
$C(\Dphi) \equiv \frac{1}{\Ncorr}\, \frac{\Delta \Npair}{\Dphi}$,
where $\Delta \Npair$ represents the number of efficiency-corrected
muon pairs in a given \Dphi interval
and \Ncorr represents the \Dphi-integrated
total number of muon pairs.
The correlation functions are constructed using 32 equal \Dphi
intervals spanning the range $[-\pi/2,3\pi/2]$.
Figure~\ref{fig:fit_PbPb_tight} shows results obtained for the
\PbPb\ 0--10\% centrality interval (top), as an example, and for the \pp\ data set (bottom).
The correlation functions are symmetrized about $\Dphi=0$,
with proper accounting of the statistical uncertainties.
\begin{figure}[h]
\centering
\includegraphics[width=1.0\textwidth]{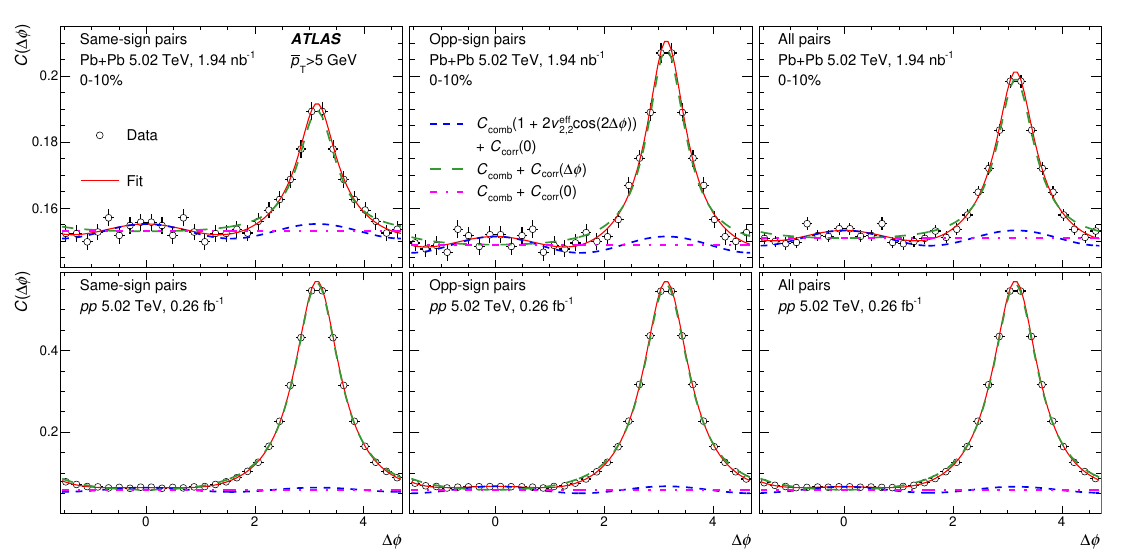}
\caption{
Measured two-muon correlation functions, $C(\Dphi)$.
Top row: results for \PbPb\ data in the 0--10\% centrality interval;
Bottom row: \pp\ results.
Left column: same-sign pairs, Middle column: opposite-sign pairs, Right column: all pairs.
Also shown are the results of fits of the correlation functions
to Eq.~\eqref{eq:fit_func}, along with different components
of the fits.
\label{fig:fit_PbPb_tight}
}
\end{figure}
All the correlation functions show a clear enhancement near $\Dphi =\pi$
superimposed on a pedestal.
That pedestal is significantly enhanced in \PbPb\ collisions due to the geometric
enhancement of hard-scattering processes that can yield multiple
uncorrelated \ccbar\ or \bbbar\ pairs in the same collision.
Moreover, in \PbPb\ collisions, this combinatoric contribution is modified by
the collective expansion of the QGP~\cite{HION-2015-06,HION-2019-11} with the result
that the combinatoric background exhibits an elliptic ($\cos{2\Dphi}$) modulation
that is evident in Figure~\ref{fig:fit_PbPb_tight}.

To characterize the shape of the peak near $\Dphi = \pi$, the data are
fit with the functional form
\begin{eqnarray}
C^{\mathrm{Fit}}(\Dphi)=C_{\text{comb}}\left[1+2\vtt\cos{\left(2\Dphi\right)}\right]+ C_{\text{corr}}(\Dphi),
\label{eq:fit_func}
\end{eqnarray}
with
\begin{eqnarray}
\Ccorr(\Dphi) = \frac{\Ccorrmax\, \Gamma^2}{(\Dphi-\pi)^2
+ \Gamma^2}.
\label{eq:fit_func2}
\end{eqnarray}
Here, $C_{\text{comb}}$ represents the combinatoric contribution to $C(\Dphi)$,
\vtt represents the Fourier coefficient for the elliptic modulation~\cite{v22ref}
of the combinatoric contribution, and \Ccorr represents the correlated
contribution, which is centered at $\Dphi=\pi$, and is parameterized
using a Cauchy-Lorentz distribution specified in Eq.~\eqref{eq:fit_func2}.
\Ccorrmax represents the maximum value of \Ccorr and \Gam the half-width at half-maximum of the
Cauchy-Lorentz distribution.
Here, $C_{\text{comb}}$, \vtt, \Ccorrmax, and \Gam, are fit parameters.
$\Ccorr(\Dphi)$ is folded at $\Dphi = -\pi/2$ and $3\pi/2$
to make it periodic in \Dphi.
Due to the long tails of the Cauchy-Lorentz distribution and the folding,
there is a ``pedestal'' to \Ccorr, separate from the $C_{\text{comb}}$ term in Eq.~\eqref{eq:fit_func},
that nonetheless represents correlated production of muon pairs.
The integrals of \Ccorr\ for the opposite-sign pairs are found to be
approximately twice of that for the same-sign pairs,
similar to expectations from \pythia.

Results of fits of Eq.~\eqref{eq:fit_func} to the
measured correlation functions are shown in Figure~\ref{fig:fit_PbPb_tight}.
Also show on Figure~\ref{fig:fit_PbPb_tight} are the individual
components of the fit:
the combinatorial pedestal (pink),
the flow modulation (blue),
and the signal component (green).
The peaks at $\Dphi \sim \pi$ are well-reproduced by the assumed form for \Ccorr.
In contrast, Gaussian, generalized-Gaussian or von-Mises forms~\cite{ALICE:2021kpy,ALICE:2023kjg}
for \Ccorr fail to describe the shape of the away-side correlation.
The modulation of the combinatoric contribution in the data is well-described
by the assumed  $\cos{\left(2\Dphi\right)}$ dependence.
In the \PbPb\ data, the \vtt values are $\mathcal{O}(0.01)$ and are
equal, within uncertainties, for same-sign and opposite-sign pairs.
This result is consistent with the hypothesis that the
modulation predominantly arises from an elliptic flow of muons
produced in  uncorrelated HF decays.
While previous measurements~\cite{HION-2018-23} have demonstrated elliptic modulation of
HF yields in \pp collisions, the \vtt value in \pp\
collisions measured here changes slightly
if the \absDy\ requirement is increased from 0.8 indicating a
residual near-side correlation that also contributes to the \vtt.
The change in the \Gam values
when the \pp\ data are fit constraining
\vtt to zero, are included in the systematic uncertainties.

\sloppy
The standard deviation of \Ccorr\ obtained from the fits,
around $\Dphi=\pi$ is evaluated as:
$\sigDphi \equiv \sqrt{\int \left(\Dphi - \pi\right)^2 \left(\Ccorr(\Dphi)-\Ccorr(0)\right)\dd{\Dphi}}$,
where the integral is performed over the interval $[0,2\pi]$.
When evaluating \sigDphi, the pedestal of $\Ccorr(\Dphi)$
i.e., $\Ccorr(0)$, is subtracted out.
Statistical uncertainties on the extracted \Gam and \sigDphi values are evaluated
using re-samplings of the measured correlation functions assuming
Gaussian-distributed statistical uncertainties for each point.
Each of the re-sampled correlations functions is fit to Eq.~\eqref{eq:fit_func}
and the standard deviation of the resulting \Gam and \sigDphi
distributions are taken as the statistical uncertainties.

Systematic uncertainties on the \Gam or \sigDphi
may arise from the muon selection,
the trigger and reconstruction efficiency corrections,
corrections for the mass selection on opposite-sign pairs,
parameterization of the combinatorial background in Eq.~\eqref{eq:fit_func},
and from the method used to extract the widths of the away-side peak.
Effects from the $\pt$, $\eta$ and $\phi$ resolution of the
reconstructed tracks are negligible in this measurement,
and not included in the systematic uncertainties.
The effect of the muon selection is studied by using the
\textit{medium} working point~\cite{PERF-2015-10}
that is less pure and, thus, has a larger contamination
from non-HF hadron (mostly pion and kaon) decays.
The effects of the trigger and reconstruction efficiency uncertainties on
the measurement are determined by varying the efficiencies
systematically up or down within their uncertainties
and repeating the analysis.
For the corrections related to the dimuon photoproduction and mass restrictions
placed on the opposite-sign pairs, the systematic uncertainty is obtained by using
an alternate estimate of $A^{\text{opp}}(\Dphi)$ from mixed-events,
as described before.
The sensitivity of the results to the parameterization of
the modulation of the combinatoric background is done as follows:
for the \PbPb\ measurements, this is done by including a $\vthth\cos{\left(3\Dphi\right)}$
term in the square brackets in Eq.~\eqref{eq:fit_func}, with \vthth\ being an additional fit parameter;
for the \pp\ measurements, for reasons stated before, the data were instead fit
constraining \vtt to zero.
This is the leading systematic uncertainty for both \sigDphi\ and $\Gamma$.
This uncertainty is $\sim1\%$ ($2.5\%$) for \sigDphi\
and $\sim2\%$ ($5.5\%$) for $\Gamma$
in the \PbPb(\pp) measurements.
The correction for the DY background pairs is evaluated
using a \powheg MC sample that uses the
nNNPDF2.0~\cite{AbdulKhalek:2020yuc} nuclear PDFs.
Systematic uncertainties are evaluated using the nCTEQ15~\cite{Kovarik:2015cma}
and NNPDF3.0~\cite{Ball_2015} PDFs.
To evaluate the sensitivity of the \sigDphi results to the assumed form
for \Ccorr, an alternative method for extracting the width is applied.
Namely, the measured correlation function over the range
$[-\pi/2,\pi/2]$ is shifted by $\pi$ and subtracted from
that measured in the range $[\pi/2,3\pi/2]$.
Then, the standard deviation, \sigDphiP, is obtained directly from the
subtracted correlation function,
with \Dphi\ restricted to $[\pi/2,3\pi/2]$.
This method produces a biased estimate of \sigDphi,
as it assumes that the away-side correlation is restricted to $[\pi/2,3\pi/2]$,
while the observed away-side peak extends beyond this range.
However, the results can be compared to a calculation of \sigDphiP from
\Ccorr using the same shifted-subtraction method.
The relative difference between the values of \sigDphiP\
-- obtained directly from the measured correlation,
and obtained from \Ccorr with this modified procedure --
is taken to be the relative systematic uncertainty on \sigDphi.

\begin{figure}
\centering
\includegraphics[width=1.0\textwidth]{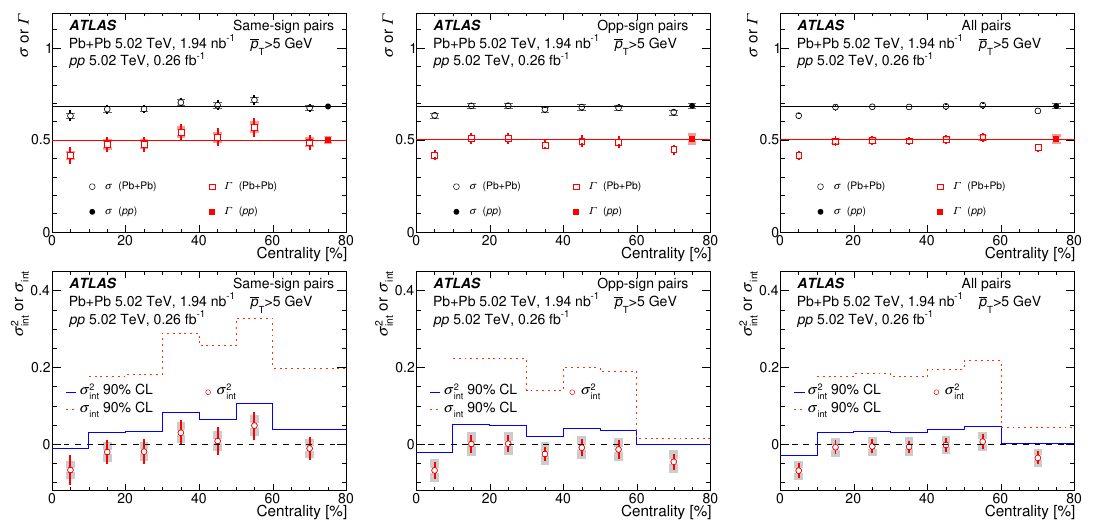}
\caption{
Top panels: The measured widths, \sigDphi, of the away-side peak in
the two-particle correlation functions in \PbPb collisions
versus centrality.
Also shown are the values of the \Gam\ parameter in Eq.~\eqref{eq:fit_func}.
The \sigDphi\ and \Gam\ measured in \pp\ collisions are also shown.
The vertical lines (bands) on the data points represent the
statistical (systematic) uncertainties.
The horizontal lines indicate the nominal \pp\ values.
Bottom panels: The difference between the squared-widths of the
\PbPb\ and \pp\ measurements.
Also shown are the upper limit on
$\sigma_{\mathrm{int}}^{2}$ and $\sigma_{\mathrm{int}}$
at the 90\% CL\@.
As the 90\% CL on $\sigma_{\mathrm{int}}^{2}$ is negative for the 0--10\%
centrality interval, the corresponding 90\% CL on
$\sigma_{\mathrm{int}}$ cannot be evaluated, and is not shown.
The left, middle, and right panels correspond to
same-sign pairs, opposite-sign pairs, and all-pairs.
\label{fig:result_widths}
}
\end{figure}

Figure~\ref{fig:result_widths} shows the measured values of \sigDphi\
and \Gam\  in \PbPb\ collisions as a function of centrality compared
to the same quantity in \pp\ collisions.
The two quantities exhibit similar behavior.
The widths are similar between same-sign and opposite-sign pairs; this
behavior is expected if \bb\ pairs dominate in the samples, as suggested by
Ref.~\cite{SupplementaryMaterial}.
No significant variation of the \sigDphi\ and \Gam\ is observed as a
function of centrality over the 10--80\% centrality range.
Over this centrality range, the \PbPb\ values are consistent with those
measured in \pp\ collisions.
A significant decrease in \sigDphi\ and \Gam\ is observed
in the 0--10\% centrality interval.
The interpretation of these results depends on the shape of the \pT\
spectra of the parent $b$ quarks~\cite{Nahrgang:2013saa}, which is
modified by the energy loss of the quarks in the plasma.
However, the shapes of the \ptbar-spectra in this measurement
are found to be similar across the different centrality
intervals, upto an overall normalization.
The $\langle\ptbar\rangle$, where the $\langle...\rangle$ indicates
averaging over all correlated pairs within
a centrality interval, does not change significantly with centrality
and is consistent with that measured in \pp\ collisions, except for
the 0--10\% centrality interval, where the $\langle\ptbar\rangle$
is $\sim3\%$ higher~\cite{meanpT}.
This change in the \ptbar\ spectra may be partially responsible for the
decrease in \sigDphi\ observed in the 0--10\% centrality interval.
However, the expectation from Ref.~\cite{Nahrgang:2013saa} of a
broadening in \Gam
that systematically
increases from peripheral to central collisions, does not appear to
be supported by the data.
A \powheg\ study done as part of this analysis, shows that over the
muon-\pt\ and $\eta$ ranges used here, the angular correlation between the
$b$-hadron ($c$-hadron) and the decay-muon has an r.m.s. width of 0.158 (0.058).
These decay-widths are significantly smaller than the angular-correlation
between the muons themselves (Figure~\ref{fig:result_widths}),
and thus the impact of the decay on the measurements is small.
The lower panels of Figure~\ref{fig:result_widths} show the squared
difference between the \PbPb\ and \pp\ widths:
$\sigma_{\mathrm{int}}^{2}=\sigma^{2}_{\mathrm{Pb+Pb}} - \sigma^{2}_{pp}$.
This quantity represents the square of the additional angular
broadening in \PbPb collisions, from interactions with the QGP~\cite{endnote1}.
Except for in the 0--10\% centrality interval, the \PbPb\ \sigDphi\ values are
typically within about one standard deviation of the \pp\ values.
Figure~\ref{fig:result_widths} also shows the upper limit of
$\sigma_{\mathrm{int}}^{2}$ (and $\sigma_{\mathrm{int}}$)
at the 90\% confidence level (CL),
evaluated using the combined statistical + systematic uncertainties.

In summary, this Letter presented results of a novel measurement of the
angular correlations between heavy-quarks produced in \pp\ and \PbPb\
collisions at the LHC using semileptonic decays of the HF hadrons to muons.
Muons with $\pt^{\mu}>4$~\GeV\ and $|\eta^\mu| < 2.4$ were used in the analysis.
Two-muon correlation functions were constructed from same-sign and
opposite-sign pairs, with \ptbar>5~\GeV, and studied as a function
of azimuthal angle difference, \Dphi, for $|\Deta|>0.8$.
A strong enhancement is observed in the correlation functions at $\Dphi \sim \pi$,
consistent with the production of the muon pairs from semileptonic decays
of heavy-quark pairs -- primarily \bb pairs -- created in hard-scattering processes.
The widths of the peaks at $\Dphi \sim \pi$, characterized by the
half-width at half-maximum (\Gam) and the standard
deviation, \sigDphi, show no significant difference between \pp\ collisions
and \PbPb\ collisions and no significant variation with centrality
except in the 0--10\% most central collisions,
where a significant decrease in the \PbPb\ widths is observed.
The results are consistent between same-sign pairs, which have negligible charm
contribution, and opposite-sign pairs for which bottom pairs
contribute $\sim$90\% of the yield in \pp collisions.
Limits, at the 90\% confidence level, are placed on the
standard deviation of additional angular deflection introduced by the QGP\@.
For the all-pairs sample,
the data limit the standard deviation of
the additional {\em pair} broadening ($\sigma_{\mathrm{int}}$)
to $\lesssim 0.2$, except in the 0--10\% interval for which the
measured narrowing of the distribution is significant at the
$\sim2\sigma$ level.
These results provide a model-independent constraint on the stochastic
deflection of bottom quarks in the QGP.


%
\section*{Acknowledgements}

%
%

%
%

We thank CERN for the very successful operation of the LHC and its injectors, as well as the support staff at
CERN and at our institutions worldwide without whom ATLAS could not be operated efficiently.

The crucial computing support from all WLCG partners is acknowledged gratefully, in particular from CERN, the ATLAS Tier-1 facilities at TRIUMF/SFU (Canada), NDGF (Denmark, Norway, Sweden), CC-IN2P3 (France), KIT/GridKA (Germany), INFN-CNAF (Italy), NL-T1 (Netherlands), PIC (Spain), RAL (UK) and BNL (USA), the Tier-2 facilities worldwide and large non-WLCG resource providers. Major contributors of computing resources are listed in Ref.~\cite{ATL-SOFT-PUB-2023-001}.

We gratefully acknowledge the support of ANPCyT, Argentina; YerPhI, Armenia; ARC, Australia; BMWFW and FWF, Austria; ANAS, Azerbaijan; CNPq and FAPESP, Brazil; NSERC, NRC and CFI, Canada; CERN; ANID, Chile; CAS, MOST and NSFC, China; Minciencias, Colombia; MEYS CR, Czech Republic; DNRF and DNSRC, Denmark; IN2P3-CNRS and CEA-DRF/IRFU, France; SRNSFG, Georgia; BMBF, HGF and MPG, Germany; GSRI, Greece; RGC and Hong Kong SAR, China; ISF and Benoziyo Center, Israel; INFN, Italy; MEXT and JSPS, Japan; CNRST, Morocco; NWO, Netherlands; RCN, Norway; MEiN, Poland; FCT, Portugal; MNE/IFA, Romania; MESTD, Serbia; MSSR, Slovakia; ARRS and MIZ\v{S}, Slovenia; DSI/NRF, South Africa; MICINN, Spain; SRC and Wallenberg Foundation, Sweden; SERI, SNSF and Cantons of Bern and Geneva, Switzerland; MOST, Taipei; TENMAK, T\"urkiye; STFC, United Kingdom; DOE and NSF, United States of America.

Individual groups and members have received support from BCKDF, CANARIE, CRC and DRAC, Canada; CERN-CZ, PRIMUS 21/SCI/017 and UNCE SCI/013, Czech Republic; COST, ERC, ERDF, Horizon 2020, ICSC-NextGenerationEU and Marie Sk{\l}odowska-Curie Actions, European Union; Investissements d'Avenir Labex, Investissements d'Avenir Idex and ANR, France; DFG and AvH Foundation, Germany; Herakleitos, Thales and Aristeia programmes co-financed by EU-ESF and the Greek NSRF, Greece; BSF-NSF and MINERVA, Israel; Norwegian Financial Mechanism 2014-2021, Norway; NCN and NAWA, Poland; La Caixa Banking Foundation, CERCA Programme Generalitat de Catalunya and PROMETEO and GenT Programmes Generalitat Valenciana, Spain; G\"{o}ran Gustafssons Stiftelse, Sweden; The Royal Society and Leverhulme Trust, United Kingdom.

In addition, individual members wish to acknowledge support from Chile: Agencia Nacional de Investigación y Desarrollo (FONDECYT 1190886, FONDECYT 1210400, FONDECYT 1230987); China: National Natural Science Foundation of China (NSFC - 12175119, NSFC 12275265); European Union: European Research Council (ERC - 948254), Horizon 2020 Framework Programme (MUCCA - CHIST-ERA-19-XAI-00), Italian Center for High Performance Computing, Big Data and Quantum Computing (ICSC, NextGenerationEU), Marie Sklodowska-Curie Actions (EU H2020 MSC IF GRANT NO 101033496); France: Agence Nationale de la Recherche (ANR-20-CE31-0013, ANR-21-CE31-0013, ANR-21-CE31-0022), Investissements d'Avenir Idex (ANR-11-LABX-0012), Investissements d'Avenir Labex (ANR-11-LABX-0012); Germany: Baden-Württemberg Stiftung (BW Stiftung-Postdoc Eliteprogramme), Deutsche Forschungsgemeinschaft (DFG - CR 312/5-1); Italy: Istituto Nazionale di Fisica Nucleare (FELLINI G.A. n. 754496, ICSC, NextGenerationEU); Japan: Japan Society for the Promotion of Science (JSPS KAKENHI 22H01227, JSPS KAKENHI JP21H05085, JSPS KAKENHI JP22H04944); Netherlands: Netherlands Organisation for Scientific Research (NWO Veni 2020 - VI.Veni.202.179); Norway: Research Council of Norway (RCN-314472); Poland: Polish National Agency for Academic Exchange (PPN/PPO/2020/1/00002/U/00001), Polish National Science Centre (NCN 2021/42/E/ST2/00350, NCN OPUS nr 2022/47/B/ST2/03059, NCN UMO-2019/34/E/ST2/00393, UMO-2020/37/B/ST2/01043, UMO-2021/40/C/ST2/00187); Slovenia: Slovenian Research Agency (ARIS grant J1-3010); Spain: BBVA Foundation (LEO22-1-603), Generalitat Valenciana (Artemisa, FEDER, IDIFEDER/2018/048), La Caixa Banking Foundation (LCF/BQ/PI20/11760025), Ministry of Science and Innovation (RYC2019-028510-I, RYC2020-030254-I), PROMETEO and GenT Programmes Generalitat Valenciana (CIDEGENT/2019/023, CIDEGENT/2019/027); Sweden: Swedish Research Council (VR 2022-03845, VR 2022-04683), Knut and Alice Wallenberg Foundation (KAW 2017.0100, KAW 2018.0157, KAW 2019.0447); Switzerland: Swiss National Science Foundation (SNSF - PCEFP2\_194658); United Kingdom: Leverhulme Trust (Leverhulme Trust RPG-2020-004); United States of America: Neubauer Family Foundation.

%
%


%

%
\clearpage
\appendix
\part*{Appendix}
\addcontentsline{toc}{part}{Appendix}
\section{Supplemental Material Allowed by Journal to Accompany Online Version of Paper}

\begin{figure}[h]
\centering
\includegraphics[width=1.0\textwidth]{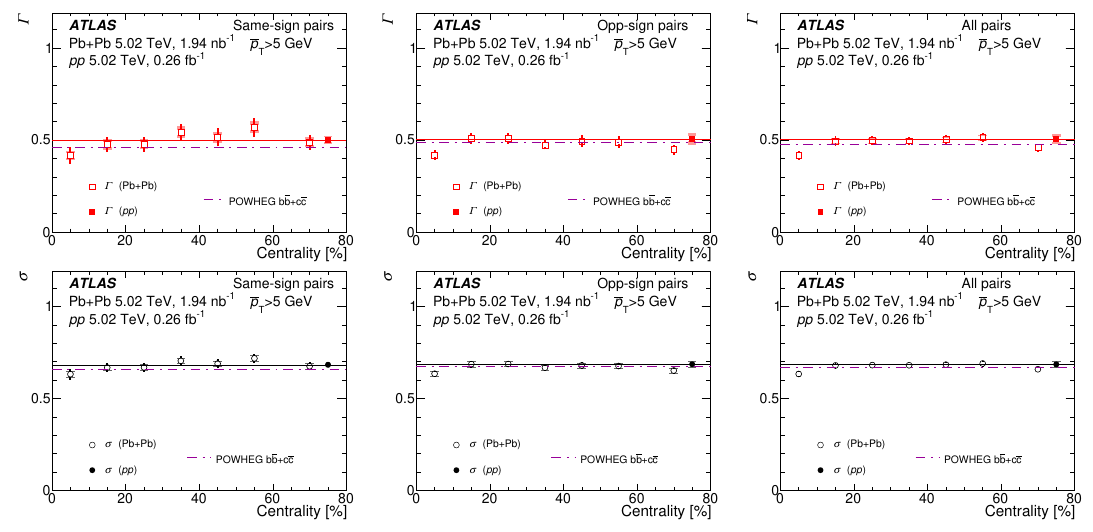}
\caption{
Comparison of the $\Gamma$ (top) and $\sigma$ (bottom) obtained from
a \powheg\ MC sample to those obtained in the data.
The horizontal continuous lines indicate the nominal \pp\ values plotted across
the full centrality range.
The \powheg\ calculation includes two samples:
a $\bbbar\rightarrow\mu\mu$ and a $\ccbar\rightarrow\mu\mu$ sample.
The same fiducial selections are applied
on the single-muons and on the muon-pair as those in the data analysis;
namely that the single-muon $|\eta|<2.4$ and $\pt>4$~\GeV, and that
for the pair, $\ptbar>5$~\GeV\ and $|\Deta|>0.8$.
The $\bbbar$ and $\ccbar$ samples are combined appropriately
weighted by the cross-sections.
The samples are generated using the nCTEQ15+SIH~\cite{PhysRevD.104.094005} nuclear PDFs
corresponding to the Pb-nucleus, and thus include the correct admixture
of $pp$, $pn$ and $nn$ collisions for comparison to the \PbPb\ measurements.
The left, center and right panels correspond to
same-sign, opposite-sign and all-pairs respectively.
\label{fig:powheg_compare1}
}
\end{figure}

\begin{figure}[h]
\centering
\includegraphics[width=1.0\textwidth]{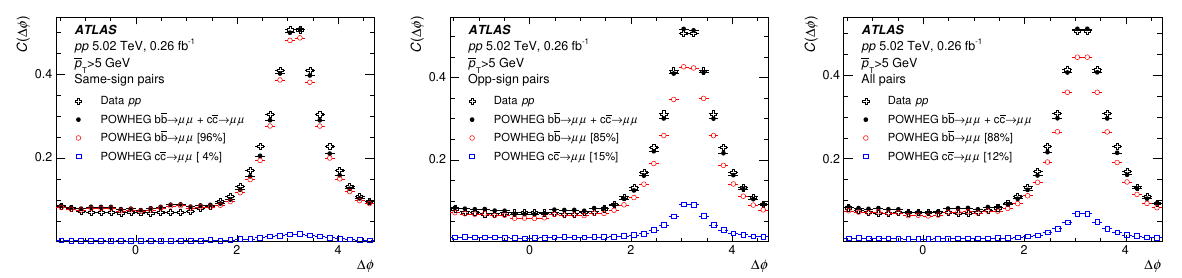}
\caption{
Correlation functions, $C(\Dphi)$, in the \powheg\ MC sample
used in Figure~\ref{fig:powheg_compare1}.
The left, center and right panels correspond to
same-sign, opposite-sign and all-pairs respectively.
The relative contribution of \bbbar\ and \ccbar\ pairs are also shown.
The measured $C(\Dphi)$ in the \pp\ data are also shown for comparison.
\label{fig:powheg_compare2}
}
\end{figure}


%

%
%

%
%
%
\printbibliography
\clearpage
 
\begin{flushleft}
\hypersetup{urlcolor=black}
{\Large The ATLAS Collaboration}

\bigskip

\AtlasOrcid[0000-0002-6665-4934]{G.~Aad}$^\textrm{\scriptsize 102}$,
\AtlasOrcid[0000-0002-5888-2734]{B.~Abbott}$^\textrm{\scriptsize 120}$,
\AtlasOrcid[0000-0002-1002-1652]{K.~Abeling}$^\textrm{\scriptsize 55}$,
\AtlasOrcid[0000-0001-5763-2760]{N.J.~Abicht}$^\textrm{\scriptsize 49}$,
\AtlasOrcid[0000-0002-8496-9294]{S.H.~Abidi}$^\textrm{\scriptsize 29}$,
\AtlasOrcid[0000-0002-9987-2292]{A.~Aboulhorma}$^\textrm{\scriptsize 35e}$,
\AtlasOrcid[0000-0001-5329-6640]{H.~Abramowicz}$^\textrm{\scriptsize 151}$,
\AtlasOrcid[0000-0002-1599-2896]{H.~Abreu}$^\textrm{\scriptsize 150}$,
\AtlasOrcid[0000-0003-0403-3697]{Y.~Abulaiti}$^\textrm{\scriptsize 117}$,
\AtlasOrcid[0000-0002-8588-9157]{B.S.~Acharya}$^\textrm{\scriptsize 69a,69b,m}$
% ...
\bigskip
\\

% ...
$^{29}$Physics Department, Brookhaven National Laboratory, Upton NY; United States of America.\\
% ...
$^{35}$$^{(a)}$Facult\'e des Sciences Ain Chock, R\'eseau Universitaire de Physique des Hautes Energies - Universit\'e Hassan II, Casablanca;$^{(b)}$Facult\'{e} des Sciences, Universit\'{e} Ibn-Tofail, K\'{e}nitra;$^{(c)}$Facult\'e des Sciences Semlalia, Universit\'e Cadi Ayyad, LPHEA-Marrakech;$^{(d)}$LPMR, Facult\'e des Sciences, Universit\'e Mohamed Premier, Oujda;$^{(e)}$Facult\'e des sciences, Universit\'e Mohammed V, Rabat;$^{(f)}$Institute of Applied Physics, Mohammed VI Polytechnic University, Ben Guerir; Morocco.\\
% ...
$^{49}$Fakult\"{a}t Physik , Technische Universit{\"a}t Dortmund, Dortmund; Germany.\\
% ...
$^{55}$II. Physikalisches Institut, Georg-August-Universit\"{a}t G\"ottingen, G\"ottingen; Germany.\\
% ...
$^{69}$$^{(a)}$INFN Gruppo Collegato di Udine, Sezione di Trieste, Udine;$^{(b)}$ICTP, Trieste;$^{(c)}$Dipartimento Politecnico di Ingegneria e Architettura, Universit\`a di Udine, Udine; Italy.\\
% ...
$^{102}$CPPM, Aix-Marseille Universit\'e, CNRS/IN2P3, Marseille; France.\\
% ...
$^{117}$Department of Physics, New York University, New York NY; United States of America.\\
% ...
$^{120}$Homer L. Dodge Department of Physics and Astronomy, University of Oklahoma, Norman OK; United States of America.\\
% ...
$^{150}$Department of Physics, Technion, Israel Institute of Technology, Haifa; Israel.\\
$^{151}$Raymond and Beverly Sackler School of Physics and Astronomy, Tel Aviv University, Tel Aviv; Israel.\\
% ...
$^{a}$ Also Affiliated with an institute covered by a cooperation agreement with CERN.\\
$^{b}$ Also at An-Najah National University, Nablus; Palestine.\\
$^{c}$ Also at Borough of Manhattan Community College, City University of New York, New York NY; United States of America.\\
$^{d}$ Also at Center for High Energy Physics, Peking University; China.\\
$^{e}$ Also at Center for Interdisciplinary Research and Innovation (CIRI-AUTH), Thessaloniki; Greece.\\
$^{f}$ Also at Centro Studi e Ricerche Enrico Fermi; Italy.\\
% ...
$^{m}$ Also at Department of Physics, King's College London, London; United Kingdom.\\
% ...

\end{flushleft}


%

%
%

%

%
%
%
%

\end{document}